\documentclass{INTERSPEECH2023}
\usepackage{graphics,hyperref,arydshln,subcaption,float,amssymb,amsfonts,caption,booktabs,algorithmic,textcomp}

\interspeechcameraready

\title{Dynamic Kernels and Channel Attention for Low Resource Speaker Verification}
\name{Anna Ollerenshaw$^1$, Md Asif Jalal$^1$, Thomas Hain$^1$}
\address{
  $^1$Speech and Hearing Group, The University of Sheffield, Sheffield, UK \thanks{This work was conducted at the Liveperson Centre of Speech and Language Technology at the University of Sheffield}}
\email{alollerenshaw1@sheffield.ac.uk}

\begin{document}

\maketitle
 
\begin{abstract}
State-of-the-art speaker verification frameworks have typically focused on developing models with increasingly deeper (more layers) and wider (number of channels) models to improve their verification performance. Instead, this paper proposes an approach to increase the model resolution capability using attention-based dynamic kernels in a convolutional neural network to adapt the model parameters to be feature-conditioned. The attention weights on the kernels are further distilled by channel attention and multi-layer feature aggregation to learn global features from speech. This approach provides an efficient solution to improving representation capacity with lower data resources. This is due to the self-adaptation to inputs of the structures of the model parameters. 
The proposed dynamic convolutional model achieved 1.62\% EER and 0.18 miniDCF on the VoxCeleb1 test set and has a 17\% relative improvement compared to the ECAPA-TDNN using the same training resources.
\end{abstract}

\section{Introduction}

\label{sec:intro}
Speaker verification (SV) aims to identify a speaker typically from an unlabeled sample of speech. This task involves measuring the similarity between a test speaker's acoustic embedding and the already enrolled target speaker embedding. This similarity is typically evaluated using distance metrics such as cosine distance or Probabilistic Linear Discriminant Analysis (PLDA). The main objective of an SV framework is to learn generalised global characteristics from speaker acoustics. Many current approaches use combinations of deep neural networks (DNNs) trained for utterance classification based upon learned features that correspond with the speaker's identity. I-vectors \cite{snyder2017deep} provide a fixed representation over the speaker acoustics; x-vectors \cite{snyder2018x} became the following state-of-the-art method for speaker representations as they were able to map variable-length utterances to embeddings of fixed dimensions. 

The topology of these models and the embedding hierarchies are hypothesised to represent different speaker characteristics. ResNet based models \cite{he2016deep} \cite{zeinali2019but} attempt to learn stronger representations with residual skip connections as this enables the composition of deeper models by learning the identity function and compensating for vanishing gradients. These speaker embeddings are further distilled by learning saliency regions with attention mechanisms. The DNN models using attention and skip connections \cite{tang2019deep} \cite{desplanques2020ecapa} proved a considerable improvement over the traditional x-vectors and i-vector embeddings. Both recurrent neural networks (RNN) and convolutional neural networks (CNN) have been used to learn temporal dependencies for speaker representations, noting that the CNN-based models have typically produced better performance with fewer number of parameters than RNNs \cite{zhao2020improving}. 

The current state-of-the-art SV architectures use Time Delay Neural Networks (TDNN) and attention mechanisms in the convolutional channel outputs, which further improved the performance results \cite{desplanques2020ecapa}. However SV is still a challenging and computationally demanding task, especially in poor acoustic conditions. Large models that have been pre-trained using huge datasets perform well \cite{chen2022large}; however, training and serving these models is becoming increasingly computationally demanding. Work by \cite{thienpondt2021integrating} introduced the CNN-ECAPA-TDNN where the convolutional front-end allows the network to construct local, frequency invariant features to integrate frequency positional information. In order to enable the network to be invariant to small shifts in the frequency domain and to compensate for the potential intra-speaker variability, 2D convolutions are used to model at a higher resolution. However, this approach also uses large amounts of training data, where typically first a pretrained large-scale model is used to then be fine-tuned for state-of-the-art results. The ResNet-based models \cite{he2016deep} can suffer from overfitting due to the increases in layer dimensionality and it can also take an excessive amount of time and computational resources to fine-tune the hyperparameters to improve the performance. Often the performance of state-of-the-art models can be difficult to replicate due to optimisation and model complexity, which has increased the uptake of fine-tuning pre-trained models \cite{vaessen2022fine, linardatos2020explainable}.

The general trend for CNN based architectures has been to increase the depth and complexity of the network, while simultaneously increasing training data size for improved accuracy \cite{simonyan2014very,szegedy2017inception}. However, considering the challenges for modelling speech data, it is becoming necessary to make systems that are more efficient with regard to size and training speed as well as more interpretable. The main contribution of this paper is to integrate attention-based dynamic kernels for convolutions for a SV task achieving similar baselines to state-of-the-art approaches with lower resources, which has not yet been explored. The proposed approach uses parallel dynamic convolutional kernels described in \ref{sec:dy_conv}, which are able to adjust parameters dependent upon the input attention. Dynamic kernels have shown promising potential for boosting the model representation capabilities without increasing the computational cost \cite{han2021dynamic} and larger models have shown performance improvements for text-independent SV \cite{kim2022temporal}. 

The proposed model builds upon the original ResNet model \cite{he2016deep}, which uses a 2D CNN based approach. This method can also be integrated into other CNN-based approaches, such as the CNN-ECAPA-TDNN for further improved SV performance without the requirement for larger or pretrained models. The main motivation for using this approach is to improve representation capacity, which is shown in the following experiments to improve verification performance without increasing the computation. This is possible with the dynamic convolution approach as the kernels share the output channels, and it is observed to outperform similar models with increased layers, parameters and training data. Section \ref{sec:results} discusses the results of the experimental models with additional details regarding the average computation time of each epoch.


\section{Model Topology}
\label{sec:pagestyle}

\subsection{Related Works}

The x-vector model, described in \cite{snyder2018x}, was developed to replace the original text-independent i-vector method, using a fully connected DNN to capture long-term dependencies to aid speaker discrimination. The architecture uses pooling layers that aggregate over the frame-level representations. The input features are spliced across the first few layers of the DNN. As the input is variable length, statistics pooling layer computes the mean and standard deviation of the frame-level representations. A PLDA backend, separately trained, is used to compare embedded pairs.

ResNet-34 models such as \cite{zhang2021duality} and \cite{chung2018voxceleb2} contain 4 residual blocks between a frame-level representation extraction module and an utterance-level aggregator. The outputs of the residual blocks are fed to attention modules which attempts to learn the channel dependencies prior to the skip connections. Frame-level speaker representations are encoded into fixed-length utterance level representations by the aggregator and classified with a softmax layer. 



\subsection{Dynamic Convolution Kernels}
\label{sec:dy_conv}

\begin{figure}[ht]
    \centering
    \includegraphics[height=0.75\linewidth,width=\linewidth]{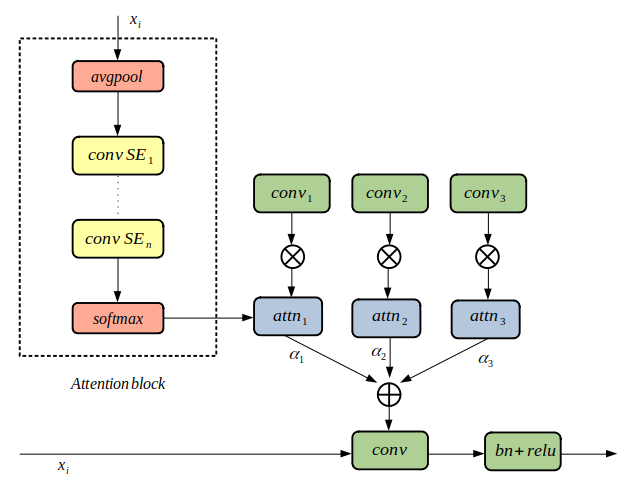}
    \caption{Dynamic kernel convolution block \cite{chen2020dynamic}, where $\alpha_k$ refers to attention weights for the $k^{th}$ linear function}
    \label{fig:dy_conv_arch}
\end{figure}

\begin{figure}[ht]
    \centering
    \includegraphics[height=0.75\linewidth,width=\linewidth]{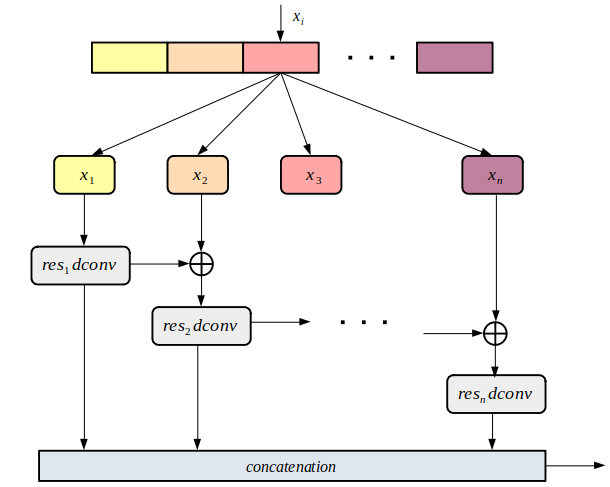}
    \caption{Residual structure of dynamic convolution (\textit{dconv}) blocks showing connections between blocks}
    \label{fig:res_arch}
\end{figure}

The dynamic convolution approach proposed in \cite{chen2020dynamic}, is a technique developed to increase the model resolution capability without requirements for increasing the model depth or width to improve accuracy. \textit{Dynamic} in this case refers to the combination of kernels for the changing input sequence by the application of input dependent attention weighting. This is integrated by aggregating parallel convolution kernels based on their attention weights, illustrated in Figure \ref{fig:dy_conv_arch}. The residual dynamic convolution block architecture is shown in Figure \ref{fig:res_arch}. Using a dynamic approach, models from other domains, such as image recognition, have been shown to have greater feature representation capacity for image classification and human pose estimation, while also being more computationally efficient due to the kernels sharing the output channels compared to the typical static convolutional models.

\begin{figure*}[ht!]
    \centering
    \includegraphics[height=0.3\linewidth,width=\textwidth]{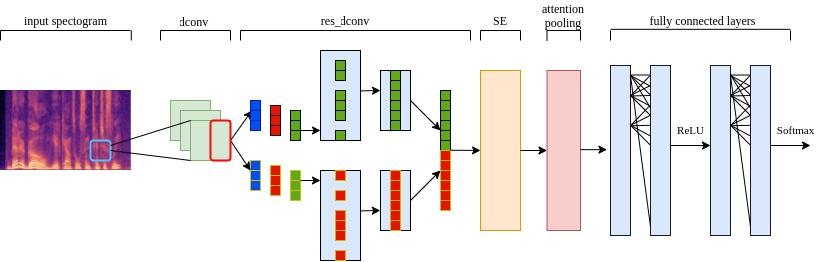}
    \caption{Model pipeline of the proposed \textit{dconv} network.}
    \label{fig:att_mech}
\end{figure*}

Equation \ref{eq:1a} describes how the input sequence \textbf{x} is utilised to compute the attention weights $\alpha$ independently within the convolution. The computation of the weight \textbf{W} matrices and bias \textbf{b} matrix are defined as the aggregation of linear functions passed through a non-linear activation function \textit{f} (ReLU in this case):

\begin{equation}
    \textbf{y}=f(\tilde{\textbf{W}}^T(\textbf{x})x + \tilde{\textbf{b}}(\textbf{x}))
    \label{eq:1a}
\end{equation}
where:
\begin{equation}
    \tilde{\textbf{W}}(x)=\sum^{K}_{k=1} \alpha_k(x) \tilde{\textbf{W}}_k 
    \label{eq:2a}
\end{equation}
and:
\begin{equation}
    \tilde{\textbf{b}}(x)=\sum^K_{k=1} \alpha_k(x)\tilde{\textbf{b}}_k
    \label{eq:3a}
\end{equation}
In Equations \ref{eq:2a} and \ref{eq:3a}, $\alpha_k$ refers to the attention weight for the $k^{th}$ linear function where $0 \leq\alpha_k(x) \leq 1, \sum^K_{k=1} \alpha_k (x) =1$. Squeeze and excitation \cite{hu2018squeeze} is applied to compute the kernel attentions, where the global spatial information is squeezed by average pooling. Due to small parallel kernel sizes and shared output channels, the aggregation of convolution kernels is computationally efficient. As joint optimisation is required for all kernels and the attention model, through multiple layers; the fully connected layers use a ReLU function and then are normalised with a softmax function for $K$ convolution kernels. Finally, after the attention weights are compiled, batch normalisation is used on the output, followed by the final ReLU function. The dynamic convolution block is denoted by the term \textit{dconv} in this paper and the \textit{dconv} block architecture is shown in Figure \ref{fig:dy_conv_arch}. The overall pipeline is illustrated in Figure \ref{fig:att_mech}, where the \textit{res\_dconv} demonstrates the residual dconv mechanism.

\subsection{Hierarchical Feature Aggregation Method for Speaker Verification}
\label{sec:hier}

Building upon the original x-vector model, the ECAPA-TDNN \cite{desplanques2020ecapa} used hierarchically grouped convolutions rather than 1-dimensional convolutions. In this paper, embeddings from multi-layer dynamic kernel convolutions are concatenated, as shown in Figure \ref{fig:res_arch}. The input features are separately chunked and fed into hierarchical layers connected with skip connections before passing to the squeeze and excitation blocks. The main motivation for using a skip connection is that is a method of compensation to collect information from previous layers and thereby the features learned by the current layer. This enables the construction of models without the need to increase the model size (number of layers, dimensionality of layers) as the model should be able to learn saliency regions. The ResNet architecture proposed in \cite{he2016deep} was inspired by VGG nets \cite{simonyan2014very} and consists of multiple stacked convolutional layers with residual connections. The representation capacity of this ResNet architecture is determined by the model width and depth (number of layers). 

Squeeze and excitation blocks were used in \cite{chen2020dynamic} which adjust the context bound frame-level features per channel over time according to the global utterance properties. The following pooling layer uses channel-and-context-dependant self-attention to attend to different speaker characteristics at different time steps for each feature map. The weighted standard deviation of the channel $\tilde{\textbf{C}}$ is shown in Equation \ref{eq:4}. The output of the attention pooling layer is a concatenation of the weights $\tilde{\textbf{W}}$ and the standard deviation.

\begin{equation}
    \tilde{\textbf{C}} = \sqrt{\sum_{k=1}^K \alpha_k(\textbf{x}) \tilde{\textbf{b}}_k^2-\tilde{\textbf{W}}(\textbf{x})^2}
    \label{eq:4}
\end{equation}

The proposed implementation of the architecture is described in Section \ref{sec:imp}.

\section{Experiments}
\label{sec:exps}

To map the voice spectograms into compact embeddings for computation, relatively shallow \textit{dconv} models were constructed. Models were built with varying layer dimensions and depth, shown in Table \ref{tab:arch2}, to control the parameter computation and observe the impact upon verification performance, shown in Table \ref{tab:vox_exps}. The training data for each model was augmented, as described in Section \ref{sec:dataaug}, as this has been shown to create sparsity and attempts to improve generalisation. The cosine distance between the vectors in the embedding space is used to measure the similarity scores.

\subsection{Dataset}

The development set and evaluation set for VoxCeleb \cite{nagrani2017voxceleb} contain both monoaural multi-speaker recordings taken from professionally edited Youtube videos, and general conversation. There are numerous challenging aspects that affect recognition within the dataset such as overlapping speech, background noise, music, laughter, applause and singing. The training and development set is split into 1,211 speakers and the testing set is split into 40 speakers for VoxCeleb1-O. The total number of utterances is 153,516 with 116 per speaker on average. Comparative cited research for SV tasks also use the considerably larger VoxCeleb2 dataset \cite{chung2018voxceleb2} for training, which contains 6112 speakers and a total of 1,128,246 utterances, however these training and serving models with this data is computationally expensive, as explored in Section \ref{sec:results}.


\subsection{Data Augmentation}\label{sec:dataaug}

After showing promising results in \cite{yamamoto2019speaker}, frequency and time domain data augmentation was performed for all models using Specaugment \cite{park2019specaugment}. This was used to attempt to increase the amount of diversity in the training data and improve model generalisation. Reverberation was convoluted with the original speech using the RIRs from \cite{habets2006room}. The augmentations were randomly chosen between babble, music, noise and reverberation.

\subsection{Implementation Details}
\label{sec:imp}

All models were trained with mean normalised 80-dimensional log Mel filterbank coefficients obtained with a 25ms window and 10ms frame shift. The spectograms were normalised by mean and variance on the frequency axis.  Random 3 second segments were taken as mini-batches to form an input dimension of 80 $\times$ 300.


The models were implemented using the PyTorch-based Speechbrain framework \cite{ravanelli2021speechbrain} and run on 1 NVIDIA RTX3060 GPU over 10 epochs. The main model pipeline is shown in Figure \ref{fig:att_mech}. The details of the model compositions are described in Table \ref{tab:arch2}, where variations of the \textit{dconv} models were built with two size dimensions of layers; 1024 and 512 dimensions, at depths of 3, 4 and a version with 5 layers. The number of attention channels for all models remained at 128 and 192 linear neurons. Adam optimisation was used to initialise the network parameters. Each setup's learning rate was set at $0.01 \times 10^{-6}$ with a learning decay of $2 \times 10^{-6}$ up to a value of 0.001. The similarity scores were verified using the cosine distance metric.

\begin{table}[t]
  \caption{Experimental architecture setup of \textit{dconv} model implementations}
  \label{tab:arch2}
  \centering
  \begin{tabular}{c c c c}
  \toprule
  \multicolumn{1}{c}{\textbf{Number of Layers}} & 
                                         \multicolumn{1}{c}{\textbf{Channels}} & \multicolumn{1}{c}{\textbf{Kernel}} & \multicolumn{1}{c}{\textbf{Dilation}} \\
    \midrule
    3 & 512 & 5,3,1 & 1,2,1 \\
    3 & 1024 & 5,3,1 & 1,2,1 \\
    4 & 512 & 5,3,3,1 & 1,2,3,1 \\
    4 & 1024 & 5,3,3,1 & 1,2,3,1 \\
    4 & 1024 & 5,3,3,3,1 & 1,2,3,4,1 \\
    5 & 1024 & 5,3,3,3,3,1 & 1,2,3,4,5,1 \\
    \bottomrule
  \end{tabular}
\end{table}

\subsection{Results and Discussion}
\label{sec:results}

Contrary to the typical scenario where the models are trained using large amounts of data, commonly with both VoxCeleb1 and 2, and extensive computational resources, here all except one of the \textit{dconv} models were trained only using VoxCeleb1 train data and 1 GPU. The motivation for this implementation is due to computational constraints as the models trained with both datasets take over 33 hours per epoch to compute. The \textit{dconv} models trained with only VoxCeleb1 achieved better performance compared to several models trained with both datasets, as shown in Table \ref{tab:vox_exps}. The x-vector model from \cite{snyder2018x} and ECAPA-TDNN from \cite{desplanques2020ecapa} were trained and evaluated using the same pipeline in Speechbrain \cite{ravanelli2021speechbrain} to represent baselines to compare the developed models. It should be noted that none of the models have had their hyperparameters optimised but are accurately repeatable using the specified resources. All the dynamic kernel based convolution models outperformed the x-vector and cited ResNet models despite not having hyperparameters tuned for optimum performance, as is commonplace for training ResNet-style architectures. The results also suggest that reducing the depth of the convolutions but widening the dimensionality of the layers improves the verification performance of the convolutional network as the 3 layer model with 1024 dimension layers achieved an Equal Error Rate (EER) of 2.89\% with miniDCF 0.275. The 4 layer \textit{dconv} model trained with both VoxCeleb1 and 2 achieves an EER of 1.62\% with miniDCF 0.18, while the 3 layer \textit{dconv} model reaches 1.64\% EER, which is an overall 17\% relative improvement to the ECAPA-TDNN model with fewer parameters. 

\begin{table}[t]
  \caption{Experimental results of models trained using VoxCeleb1 and VoxCeleb2, evaluated on the VoxCeleb1-O test set}
  \label{tab:vox_exps}
  \centering
  \begin{tabular}{c c c c c }
  \toprule
  \multicolumn{1}{c}{} & \multicolumn{1}{c}{\textbf{Vox1-O}} & \multicolumn{2}{c}{\textbf{Training set}} & \multicolumn{1}{c}{} \\
  \multicolumn{1}{c}{\textbf{Model}} & 
                                         \multicolumn{1}{c}{\textbf{EER\% }} &
                                         \multicolumn{1}{c}{\textbf{Vox 1}} &
                                         \multicolumn{1}{c}{\textbf{Vox 2}} &\multicolumn{1}{c}{\textbf{Params}}\\
    \midrule
    ResNet-34 \cite{xie2019utterance} & 10.48 & \checkmark & \checkmark & 10m \\
    VGG-M \cite{nagrani2017voxceleb} & 10.2 & \checkmark & x & 67m \\
    ResNet-34 \cite{chung2018voxceleb2} & 5.04 & \checkmark & \checkmark & 63.5m \\
    X-vector \cite{snyder2017deep} & 4.33 & \checkmark & x & 8.2m \\
    ECAPA-TDNN \cite{desplanques2020ecapa}& 1.95 & \checkmark & \checkmark & 22.2m \\
    \hline
    Dconv-4 (1024) & \textbf{1.62} & \checkmark & \checkmark & 21m \\
    Dconv-3 (1024) & 1.64 & \checkmark & \checkmark & 12.1m \\
    \hline
    Dconv-5 (1024) & 2.946 & \checkmark & x & 32m\\
    Dconv-4 (1024) & 2.926 & \checkmark & x & 21m \\
    Dconv-4 (512) & 2.935 & \checkmark & x & 6.4m \\
    Dconv-3 (1024) & \textbf{2.89} & \checkmark & x & 12.1m \\
    Dconv-3 (512) & 2.941 & \checkmark & x & 3.9m \\ 
    \bottomrule
  \end{tabular}
\end{table}

The x-vector model has a worse performance despite attaining lower validation loss than the other models, suggesting there is poor generalisation capability within this model. Another key observation from the results, is that the 3 layer and 4 layer \textit{dconv} models that have a dimension of 1024 perform similarly, which suggests that within the structure of the embeddings compiled by the dynamic kernels, critical context is learned and contained across the dimensionality of the layers rather than across the depth (number of layers) of the models. Despite the reduced parameters of the 3 layer (12 million parameters) model compared to the 4 layer (21 million parameters) model, it is possible to retain the modelling accuracy using the proposed approach by improving the embedding representation capabilities. 

Table \ref{tab:time} displays the average computation across the \textit{dconv} models and the X-vector baseline using an NVIDIA RTX3060 GPU. The average computation time for an epoch using the X-vector approach took approximately 2 hours, which is slightly faster than an epoch for the \textit{dconv} models, however to achieve the EER performance of 4.33\%, the number of epochs was increased to 25. The number of epochs for all \textit{dconv} models was 10 to attain the results listed in Table \ref{tab:vox_exps}, therefore while each epoch with a 3 or 4 layer \textit{dconv} model may take longer to compute, the models will finish training with less overall time and with a slightly improved performance. To train the \textit{dconv} model on both VoxCeleb datasets took an average of 33 hours per epoch, while the ECAPA-TDNN model took an average of 24 hours per epoch.

\begin{table}[t]
  \caption{Average epoch computation time of models training on VoxCeleb1 using an NVIDIA RTX3060 GPU}
  \label{tab:time}
  \centering
  \begin{tabular}{c c}
  \toprule
  \multicolumn{1}{c}{\textbf{Layer Name}} & 
                                         \multicolumn{1}{c}{\textbf{Average computation time per epoch}} \\
    \midrule
    X-vector \cite{snyder2017deep} & 02:05:12 \\
    ECAPA-TDNN \cite{desplanques2020ecapa} & 03:04:20 \\
    Dconv-3 (512) & 02:46:43 \\
    Dconv-3 (1024) & 03:45:46 \\
    Dconv-4 (512) & 03:23:27 \\
    Dconv-4 (1024) & 05:47:35 \\
    Dconv-5 (1024) & 08:33:18 \\
    \bottomrule
  \end{tabular}
\end{table}



\section{Conclusion}
\label{sec:conc}

A novel approach for speaker verification has been proposed that provides improved representational capabilities while controlling network dimensionality, allowing the use of lower resources for training and computation. The \textit{dconv} model can be trained to extract high resolution features while being computationally inexpensive. Several iterations of the \textit{dconv} model were evaluated on VoxCeleb 1 and compared to a baseline x-vector model, which demonstrated the proposed approach's effectiveness at lowering the EER with low resources. It was observed for the task of speaker verification, dynamic convolutional spatial dimensions (width) contribute to a slightly increased performance improvement than increasing model depth (layerwise).
This work could be further extended across different variations of architectures and also for other domains such as speech recognition or diarisation.


\bibliographystyle{IEEEtran}
\bibliography{mybib}

\end{document}